\documentclass[11pt,a4paper]{article}

\usepackage[dvips]{graphics}
\usepackage{graphicx}
\usepackage{afterpage}
\usepackage{enumerate}
 \usepackage[cp850]{inputenc}
\newcount\longrefs
\def\aap{\ifnum\longrefs=0 {\it Astron.\ Astrophys.}\else
{A\hbox{\rm \&}A}\fi}
\def\aapr{\ifnum\longrefs=0 {Astron.\ Astrophys.\ Rev.}\else
{A\hbox{\rm \&}AR}\fi}
\def\aaps{\ifnum\longrefs=0 {\it Astron.\ Astrophys.\ Suppl.}\else
{A\hbox{\rm \&}A Suppl.}\fi}
\def\aj{\ifnum\longrefs=0 {\it Astron.\ J.}\else
{AJ}\fi}
\def\ao{\ifnum\longrefs=0 {Applied Optics}\else
{Appl.\ Opt.}\fi}
\def\aspcs{\ifnum\longrefs=0 {\it Astron.\ Soc.\ Pacific Conf. Series}\else
{ASP Conf.\ Ser.}\fi}
\def\apj{\ifnum\longrefs=0 {\it Astrophys.\ J.}\else
{ApJ}\fi}
\def\apjl{\ifnum\longrefs=0 {Astrophys.\ J. Lett.}\else
{ApJ}\fi}
\def\aplett{\ifnum\longrefs=0 {Astrophys.\ J. Lett.}\else
{ApJ}\fi}
\def\apjs{\ifnum\longrefs=0 {\it Astrophys.\ J. Suppl.}\else
{ApJS}\fi}
\def\apss{\ifnum\longrefs=0 {Astrophys.\ and Space Science}\else
{Astrophys.\ Space Sci.}\fi}
\def\araa{\ifnum\longrefs=0 {Ann.\ Rev.\ Astron.\ Astrophys.}\else
{ARA\hbox{\rm \&}A}\fi}
\def\azh{\ifnum\longrefs=0 {\it Astronomicheskii Zhurnal}\else
{Astron.\ Zhur.}\fi}
\def\baas{\ifnum\longrefs=0 {Bull.\ Am.\ Astron.\ Soc.}\else
{BAAS}\fi}
\def\bain{\ifnum\longrefs=0 {Bull.\ Astronom.\ Institutes Netherlands}\else
{Bull.\ Astr.\ Inst.\ Neth.}\fi}
\def\gca{\ifnum\longrefs=0 {Geochim.\ Cosmochim.\ Acta}\else
{Geochim.\ Cosmochim.\ Acta}\fi}
\def\grl{\ifnum\longrefs=0 {Geophys.\ Res.\ Lett.}\else
{Geoph.\ Res.\ Lett.}\fi}
\def\iaucirc{\ifnum\longrefs=0 {IAU Circulars}\else
{IAU Circ.}\fi}
\def\ip{\ifnum\longrefs=0 {in press}\else
{in press}\fi}
\def\jgr{\ifnum\longrefs=0 {J.\ Geophys.\ Res.}\else
{J.\ Geophys.\ Res.}\fi}
\def\jrasc{\ifnum\longrefs=0 {J.\ Royal Astron.\ Soc.\ Canada}\else
{JRAS Can.}\fi}
\def\mnras{\ifnum\longrefs=0 {\textit {Mon.\ Not.\ Roy.\ Astron.\ Soc.}} \else
{MNRAS}\fi}
\def\nat{\ifnum\longrefs=0 {Nature}\else
{Nat}\fi}
\def\pasj{\ifnum\longrefs=0 {\it Pub.\ Astron.\ Soc.\ Japan}\else
{PASJ}\fi}
\def\pasp{\ifnum\longrefs=0 {\it Pub.\ Astron.\ Soc.\ Pacific}\else
{PASP}\fi}
\def\physscr{\ifnum\longrefs=0 {Physica Scripta}\else
{Phys.\ Scrip.}\fi}
\def\planss{\ifnum\longrefs=0 {Planetary \& Space Science}\else
{Plan. \& Space Sci.}\fi}
\def\procspie{\ifnum\longrefs=0 {Proc.\ SPIE}\else
{Proc.\ SPIE}\fi}
\def\qjras{\ifnum\longrefs=0 {Quarterly J.\ Royal Astron.\ Soc.}\else
{QJRAS}\fi}
\def\sa{\ifnum\longrefs=0 {\it Soviet Astron..}\else
{Sov.\ Astron.}\fi}
\def\skytel{\ifnum\longrefs=0 {Sky \& Telescope}\else
{Sky \& Tel.}\fi}
\def\solphys{\ifnum\longrefs=0 {\it Solar Phys.}\else
{Sol.\ Phys.}\fi}
\def\ssr{\ifnum\longrefs=0 {\it Space Science Rev.}\else
{Space\ Sci.\ Rev.}\fi}

\begin{document}
%%==========================================================
\title{\bf Turbulence and Rotation in Solar-Type Stars}
\author{\bf V. A. Sheminova}
 \date{}

 \maketitle
 \thanks{}
\begin{center}
{Main Astronomical Observatory, National Academy of Sciences of
Ukraine,
\\Akademika  Zabolotnoho 27,  Kyiv,  03143 Ukraine\\ e-mail: shem@mao.kiev.ua}
\end{center}

 \begin{abstract}

{Microturbulence, macroturbulence, thermal motion, and rotation contribute to the broadening
of line profiles in stellar spectra. Reliable data on the velocity distribution of turbulent motions
in stellar atmospheres are needed to interpret the spectra of solar-type stars unambiguously in exoplanetary
research. Stellar spectra with a high resolution of 115000 obtained with the HARPS spectrograph
provide an opportunity to examine turbulence velocities and their depth distributions in the
photosphere of stars. Fourier analysis was performed for 17 iron lines in the spectra of 13 stars with an
effective temperature of 4900--6200 K and a logarithm of surface gravity of 3.9--5.0 as well as in the
spectrum of the Sun as a star. Models of stellar atmospheres were taken from the MARCS database.
The standard concept of isotropic Gaussian microturbulence was assumed in this study. A satisfactory
fit between the synthesized profiles of spectral lines and observational data verified the reliability of
the Fourier method. The most likely estimates of turbulence velocities, the rotation velocity, and the
iron abundance and their photospheric depth distribution profiles were obtained as a result. Microturbulence
does not vary to any significant degree with depth, while macroturbulence has a marked depth
dependence. The macroturbulence velocity increases with depth in the stellar atmosphere. The higher
the effective temperature of a star and the stronger the surface gravity, the steeper the expected 
macroturbulence gradient. The mean macroturbulence velocity increases for stars with higher temperatures,
weaker gravity, and faster rotation. The mean macro- and microturbulence velocities are correlated
with each other and with the rotation velocity in the examined stars. The ratio between the
macroturbulence velocity and the rotation velocity in solar-type stars varies from 1 (the hottest stars)
to 1.7 (the coolest stars). The age dependence of the rotation velocity is more pronounced than that of
the velocity of macroturbulent motions.
}
\end{abstract}

 \vspace {.20 cm}

%\newpage
\section{Introduction}

Main-sequence F, G, and K solar-type stars are examined in this study. Nonthermal velocities (specifically,
macroturbulence and rotation) are the primary cause of line profile broadening in the spectra of these
stars. Macroturbulence is associated with granulation, supergranulation, oscillations, and other large-scale
motions. The available measurement data
 \cite{2010MNRAS.405.1907B, %Bruntt
 2014MNRAS.444.3592D,  %Doyle 2014
 1998A&A...334..221G, %Gonzales
 1978SoPh...59..193G, % Gray 1978,
 1982ApJ...262..682G, % Gray 1982
 1984ApJ...281..719G, % Gray 1984
% 1982MmSAI..53..931G, %Gray 1982 obsor
 1997MNRAS.284..803S,  %Saar,
 1978ApJ...224..584S, %Smith 1978
 2005ApJS..159..141V}  %Valenti
suggest that macrotubrulent and rotation velocities for the indicated types of stars are comparable,
produce the same effect on the shape of spectral-line profiles, and increase with effective temperature
and luminosity. Reliable data on the changes in the macroturbulence with depth into stellar atmospheres
are lacking, it is only known that the macroturbulence velocity values determined based on weak lines are
higher than those determined for strong lines. The classical method of macroturbulence research is the
comparative analysis of line profiles synthesized and observed in the wavelength scale. If the rotation parameters
are not known beforehand, it is virtually impossible to determine macroturbulence velocities based on line
profiles. The Fourier method, which is more complex and is used less often than the classical method,
provides an opportunity to estimate them.

The Fourier method was used in the 1970s by  Gray  and Smith
\cite{1973ApJ...184..461G,
1975ApJ...202..148G,
1977ApJ...218..530G,
1978SoPh...59..193G,
1976ApJ...208..487S,
1978ApJ...224..584S,
1979PASP...91..737S}. The macrobroadening
function was initially presented as a convolution of two functions: the rotation function, which depends on
the position on the disk, and the isotropic macroturbulence function with the Gaussian model (GM). It was
demonstrated in  \cite{1975ApJ...202..148G, 1977ApJ...218..530G} that the actual macroturbulence function has 
broader wings and a narrower core
than the Gaussian function. The radial-tangential model (RTM) of velocities was proposed as an alternative,
since macroturbulence is a manifestation of the granulation velocity field. Penetrative convection shapes a
granulation pattern on the stellar surface by upward and downward convective flows and horizontal motions
between them. Therefore, macroturbulence may be approximated by two flows directed along and transversally
to the radius of a star. Experience showed that the RTM does not always provide a good fit to observational
data. Later, Gray \cite{1982ApJ...262..682G} presented a unified macrobroadening function incorporating
rotation and macroturbulence effects. This function was calculated by integrating numerically over the
stellar disk (see \cite{2005oasp.book.....G} for details). Once calculated, the macrobroadening function may
be applied to different sets of observational data. In addition, averaging over several lines, one may
derive a single solution defining the macroturbulence and rotation parameters. This approach is now used
often  \cite{2014ApJ...796...88G, 2017ApJ...845...62G, 2018ApJ...857..139G}.

Takeda \cite{2017PASJ...69...46T} recently expressed doubts regarding the applicability of RTM to solar-type
dwarf stars and presented the following arguments. The macroturbulence velocity for the Sun as a star
determined in  \cite{1977ApJ...218..530G, 2018ApJ...857..139G} ($\approx$4~km/s) is significantly higher than the
typical values of convective photospheric velocities (2--3 km/s) determined directly based on high-resolution
spectroscopic data. It is also higher than the values for the center and the limb of the solar disk
\cite{1986SoPh..106..237G, 1982SoPh...78...39K, 1984AAfz...51...42S}. In order
to clarify this issue, nonthermal velocities were studied in \cite{2017PASJ...69...46T} by analyzing a large
number of profiles of spectral lines, which were obtained in high-resolution observations in different
parts of the solar disk using the procedure of profile fitting. It turned out that macroturbulence
velocities follow an almost normal distribution without any signs of the special distribution
expected in the RTM case. Takeda has concluded that the velocity field in the solar photosphere is more
chaotic than in the RTM. It was also found that the classical anisotropic model $\zeta_{\rm ma}^2 =
(\zeta_{\rm rad} * \cos \theta)^2 + (\zeta_{\rm tan} * \sin \theta)^2$,  where $\zeta_{\rm rad}$ and
$\zeta_{\rm tan} $ are macroturbulence velocities with a Gaussian distribution in radial and tangential
directions, serves as a good approximation of the macroturbulence velocity field in the solar photosphere.  
Using this model, Takeda has obtained $\zeta_{\rm rad} \approx 2$~km/s and $\zeta_{\rm tan} \approx 2.5$~km/s 
and concluded that the complex RTM is not suitable for characterizing macroturbulence in solar-type stars, 
while the classical GM is valid and convenient.

Fourier analysis with GM and RTM was performed in our previous study \cite{2017KPCB...33..217S}  
to interpret the spectra of two stars and the solar flux. The obtained data did not reveal any clear 
advantages of RTM in the context of matching the velocity models to observational data. In our view, 
the simple Gaussian model is a reasonable (and even advantageous) alternative to RTM in routine spectral 
analysis of soar-type stars.

The aim of this study is to determine the micro- and macroturbulence velocities, the rotation velocity,
and the iron abundance for 13 solar-type stars and to examine the depth profiles of turbulence velocities
and the dependences of these velocities on the fundamental parameters of stars.

%%%%%%%%%%%%%%%%%%%%%%%%%%%%%%%%
\section{Analysis of the broadening of spectral lines}
%%%%%%%%%%%%%%%%%%%%%%%%%%%%%%%%
In order to analyze the broadening of line profiles in the spectra of slowly rotating solar-type stars, we
have adapted the Fourier technique  to the case when macroturbulence velocity $\zeta$,
microturbulence velocity $\xi$, projection  $v \sin i$ of the rotation velocity, and element abundance $A$ are 
the unknown parameters (see \cite{2017KPCB...33..217S}). Let us outline the key stages of analysis.

It was assumed that thermal function $H(\lambda)$ is independent of the position on the stellar disk, and
observed line profile $D(\lambda)$ may be represented by a double convolution:

\begin{equation}\label{Eq1}
D(\lambda)= H(\lambda)\ast M(\lambda)\ast I(\lambda).
\end{equation}
Here, $M(\lambda)$  is the macrobroadening function and $I(\lambda)$ is the instrumental broadening function. 
Asterisks denote the convolution operation. Since convolution turns into multiplication in the Fourier domain, 
the Fourier transform of the observed profile is the product of the corresponding transforms

\begin{equation}\label{Eq2}
d(\sigma) =h(\sigma) m(\sigma) i(\sigma),
\end{equation}
where  $\sigma $ [s/km] is the Fourier frequency and lower-case letters correspond to the transforms of functions 
from Eq. (1). Having divided the observed-line transform by thermal and instrumental transforms, we obtain the 
so-called residual transform:

\begin{equation}\label{Eq3}
m(\sigma)= d(\sigma)/ h(\sigma)/i(\sigma),
\end{equation}
which contains data on macrobroadening function $M(\lambda)$. In order to retrieve this data, one should define 
the model of velocities of macroturbulent motion and stellar rotation. Let us assume that the distribution of 
macroturbulence velocities is isotropic and may be represented by Gaussian function $\Theta(\lambda) $ with the 
most likely macroturbulence velocity $ \zeta$. Let us also assume that stellar rotation is of a solid-state 
nature and the rotation profile is set by the position-dependent classical rotation function $G(\lambda)$ with 
parameter  $v \sin i$.  Function $M(\lambda)$  is then a convolution of these two functions:

\begin{equation}\label{Eq2}
M(\lambda)= \Theta({\lambda})\ast G(\lambda).
\end{equation}
Varying $\zeta$  and $v \sin i$,  one may find the best fit between the transform of function $M(\lambda)$ and 
the residual transform in the noise-free frequency region and, thus, determine these two unknown parameters.

Thermal function $H(\lambda)$ is easy to calculate using the standard procedure and the atmospheric models
of stars obtained by interpolating data from the MARCS database \cite{2008A&A...486..951G}.  Effective 
temperature  $ T_{\rm eff}$, surface gravity $ \log g$, and metallicity [M/H] were taken 
from \cite{2017MNRAS.468.4151I}.  The chemical composition of the Sun agreed with the data 
from \cite{2005ASPC..336...25A}. The absorption coefficients were calculated in accordance with 
the SPANSAT algorithm \cite{1988ITF...87P....3G}, and the van der Waals damping constant was calculated 
using the Anstee-Barklem-O'Mara method  {\cite{2005A&A...435..373B, 2000A&AS..142..467B}. Since function 
$H(\lambda)$ is a convolution of the Gaussian thermal profile with the Gaussian microturbulent profile, 
it  depends on the most likely microturbulence velocity $\xi$ and element abundance $A$. The values 
of these parameters are determined by comparing the calculated equivalent width of the thermal profile and 
the observed equivalent line width.

The following iterative procedure allows one to solve this problem with four unknown parameters: (0) 
transforms $ i(\sigma )$ and $ d(\sigma )$ are calculated; (1) initial value $\xi=0.5$~km/s is set, proper 
$A$ is fitted, and  $H(\lambda)$, $ h(\sigma )$, and $ m(\sigma )$ are calculated; (2) initial value  
$V \sin i=0.5$~km/s is set and a set of functions $M(\lambda)$  and their transforms is calculated 
for $\zeta=$  0.5, 1.0, 1.5, 2.0, 2.5, 3.0~km/s; (3) these transforms are compared with residual 
transform $ m(\sigma )$ and the minimum deviation is determined; (4) operations (2) and (3) are repeated 
for other values of $V \sin i$ (1.0, 1.5, 2.0, 2.5, 3.0~km/s); (5) all operations starting from (1) are repeated 
for a different $\xi$ value (1.0, 1.5, 2.0, 2.5~km/s); (6) the smallest minimum deviation, which defines all 
four unknown parameters, is found; (7) the line profile is calculated with the obtained values of $\xi$,  
$\zeta$, $V \sin i$, and $A$ and is compared to the observed profile in the wavelength scale.

We have carefully selected a set of lines of neutral and ionized iron from the database 
\cite{1999A&AS..138..119K}  and the spectrum of the Sun as a star \cite{2005ASPC..336..321H}. This list is 
rather small but remains unchanged for all stars (Table 1). The lines within it were checked for the 
availability of a blend-free profile (at least one wing), accurate oscillator strengths, and parameters 
for calculating the damping constant. The lines were also chosen so as to maximize the range of depths of 
their formation. Their equivalent widths $W$ for the studied stars fell within the range of $260>W>20$~mA. 
Oscillator strengths $\log gf$ with an error of 3--10\% were taken from \cite{2006JPCRD..35.1669F} for
 Fe I lines and from \cite{2009A&A...497..611M} for Fe II. The spectrum of the Sun as a star with a resolution 
 of 300000 was taken from \cite{2005ASPC..336..321H}. Stellar spectroscopic data derived from the results of 
 observations \cite{2009MNRAS.398..911J} with the HARPS spectrograph mounted on the ESO 3.6 m Telescope at 
 La Silla Observatory (Chile) were provided by Ya.~Pavlenko and A.~Ivanyuk. The signal-to-noise ratio and 
 the resolving power of HARPS are higher than 100 and approximately equal to 115000, respectively.

The observed and model residual transforms were matched for each line individually so that the
obtained velocity parameters could be tied to the depth of formation of a specific line, which was calculated 
using the depression contribution function in accordance with \cite{2015arXiv150500975G}. If macroturbulence is 
assumed to be independent of depth, the residual transforms for all lines of a given star should match, 
and an averaged residual transform may be used for fitting. In reality, the residual is affected by the 
imperfect line correction due to blends, errors in observation processing, and inaccuracies in choosing 
the continuum level and calculating the thermal profile. It was demonstrated 
in \cite{2014ApJ...796...88G, 2017ApJ...845...62G, 2018ApJ...857..139G} that, despite the probable errors, the
residual transform averaged over all lines yields a reliable result. Figure 1 demonstrates the observed
symmetrized line profiles and their Fourier transforms for the Sun and HD 189627. The best fit between
the residual line transforms is achieved at the lowest frequencies. The deviation increases at medium 
frequencies ($-1.5<\log \sigma <-1$) due to the macroturbulence velocity gradient. The spread for HD~189627 
is even larger, since the corresponding spectral resolution and the signal-to-noise ratio of observations are 
lower. At higher frequencies ($\log \sigma >-1$), observation noise is also intensified due to the fact that 
the observed transform is divided by the thermal one. We have compared the solution based on the averaged
transform to the result averaged over all individual lines and obtained a satisfactory fit. Therefore, the use 
of an averaged residual transform should speed up the analysis considerably if one needs to obtain the
parameters of turbulence velocities averaged over a large number of lines.
%%%%%%%%%%%%%%%%%%%%%%%%%%%%%%%%%%%%%%%%%%% Figure 1
 \begin{figure}[t]
 \centerline{
 \includegraphics   [scale=0.9]{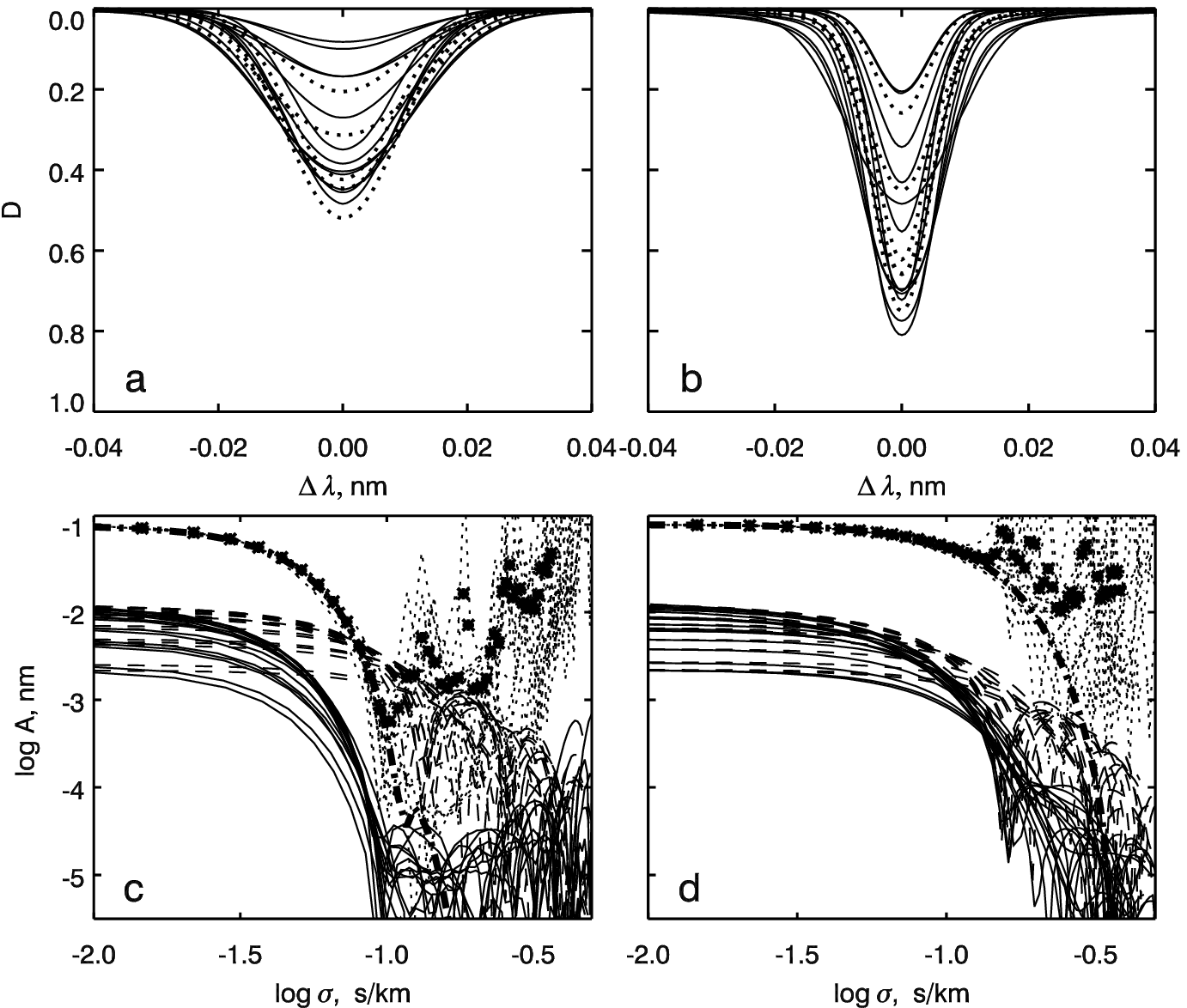} }
  %\caption
   \vspace {0.3 cm}
  {\small
Fig. 1. Profiles $D=1-F_{\lambda}/F_c$ of all the analyzed observed symmetrized Fe I (solid curves)
and Fe II (dotted curves) lines for (a) the rapidly rotating star HD 189627 and (b) the Sun and
amplitudes $\log A$ of their Fourier transforms (solid curves),
transforms of thermal profiles (dashed curves), residual transforms (dotted curves), averaged residual
transform (asterisks),
and modeled transform of the macrobroadening function (dash-and-dot curve) for (c) HD 189627 and (d) the Sun
as functions of frequency $\log \sigma $.
 }
\label{prof1}
 \end{figure}

%%%%%%%%%%%%%%%%%%%%%%%%%%%%%%%%%%%%%%%%%%%
%%%%%%%%%%%%%%%%%%%%%%%%%%%%%%%%%%%%%%%%%%% Figure 2
\begin{figure}[t]
 \centerline{
 \includegraphics   [scale=0.95]{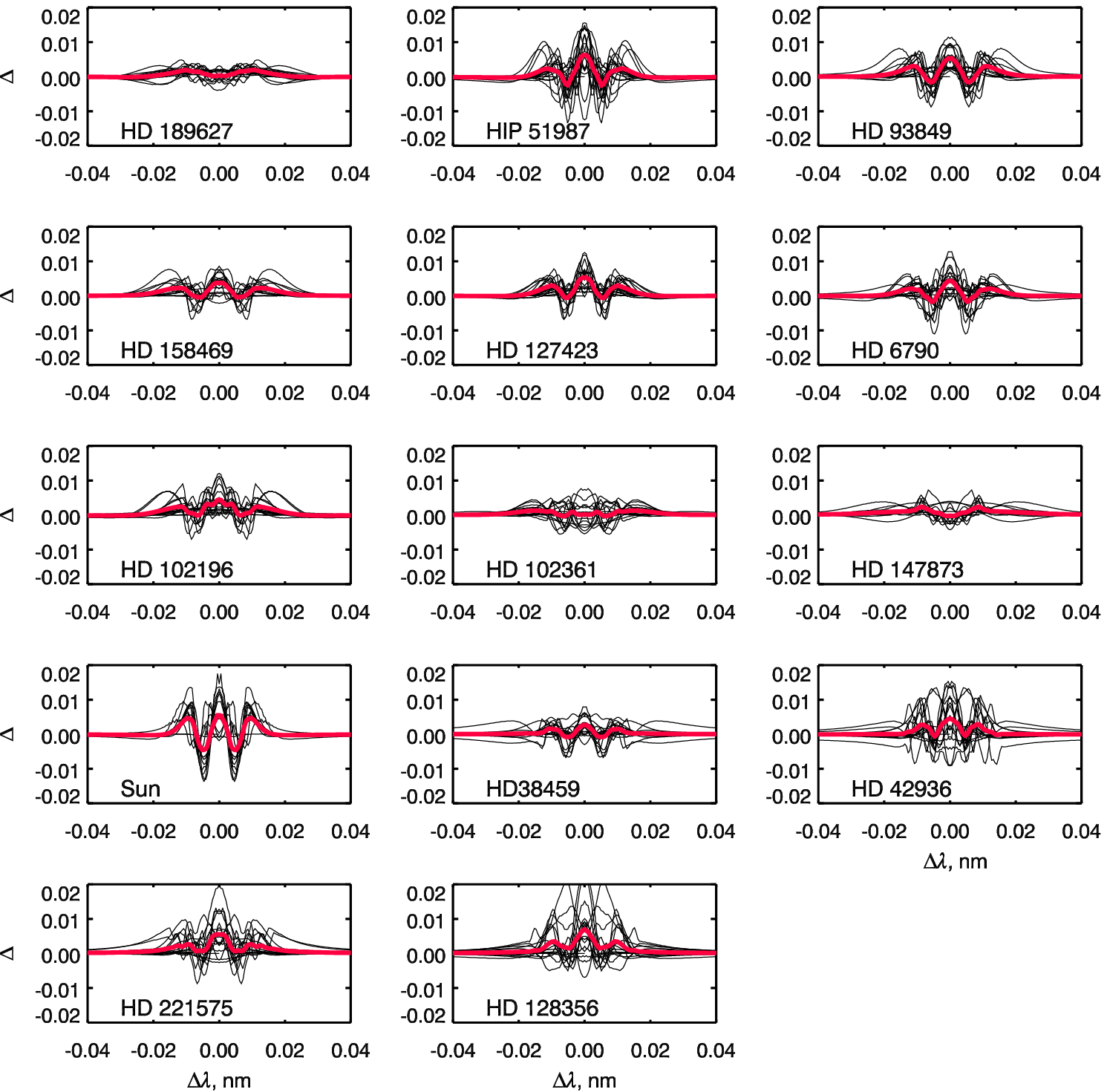}}
   %\caption
%    \vspace {0.5 cm}
   {\small
Fig. 2. Profiles of deviations $\Delta$ between the observed $D_{\rm obs}$ and synthesized $D_{\rm syn}$
line profiles as functions of distance $\Delta \lambda$ from the line center for all lines in each star.
Bold red curves represent mean profiles of deviations. Line profiles were calculated
as $ D=1-F_\lambda/ F_c$.
}
\label{prof1}
 \end{figure}

%%%%%%%%%%%%%%%%%%%%%%%%%%%%%%%%%%%%%%%%%%%

%%%%%%%%%%%%%%%%%%%%%%%%%%%%%%%%
%%==========================================================

%___________________________________ Table 1
\begin{table}
\centering
 \caption
 {\small  Parameters of the spectral lines used, their equivalent widths
 and effective formation depths for the Sun.
 } \vspace {0.3 cm}
\label{T:1}
\footnotesize{
%\small{
\begin{tabular}{ccccc}
\hline\hline
$\lambda$ & $E_{\rm exp}$   & $\log gf$ & W   &$\log \tau_5$   \\
(nm)      & (eV)            &           & (nm)&               \\
\hline
\multicolumn{5}{c} {FeI}  \\
 448.42198 & 3.603 &  -0.864 &11.40& -1.70 \\
 460.20006 & 1.608 &  -3.154 & 7.71& -2.10 \\
 499.41295 & 0.915 &  -3.080 &11.71& -2.62 \\
 524.24905 & 3.635 &  -0.968 & 9.43& -1.87 \\
 537.95734 & 3.695 &  -1.514 & 6.59& -1.59 \\
 550.14653 & 0.958 &  -3.047 &12.22& -2.67 \\
 566.13455 & 4.285 &  -1.756 & 2.46& -1.00 \\
 570.54646 & 4.302 &  -1.355 & 4.15& -1.15 \\
 577.84533 & 2.588 &  -3.430 & 2.49& -1.23 \\
 606.54848 & 2.607 &  -1.530 &12.88& -2.24 \\
 615.16170 & 2.175 &  -3.299 & 5.20& -1.63 \\
 625.25546 & 2.403 &  -1.687 &13.50& -2.32 \\
  \multicolumn{5}{c} {FeII}  \\
450.82802 & 2.860 &   -2.440 &9.53 &-1.80 \\
457.63330 & 2.840 &   -2.950 &6.98 &-1.53 \\
523.46228 & 3.220 &   -2.180 &9.18 &-1.76 \\
541.40730 & 3.223 &   -3.580 &2.97 &-0.90 \\
645.63830 & 3.904 &   -2.050 &6.72 &-1.45 \\
\hline
\end{tabular}
}
\end{table}
\noindent
%%%%%%%%%%%%%%%%%%%%%%%%%%%%%%%%%%\lhd%%%%%%%%%%%%%%%%%%
%%%%%%%%%%%%%%%%%%%%%%%%%%%%%%%%%%%%%%%%%%% Figure 3
 \begin{figure}[t]
 \centerline{
 \includegraphics   [scale=0.9]{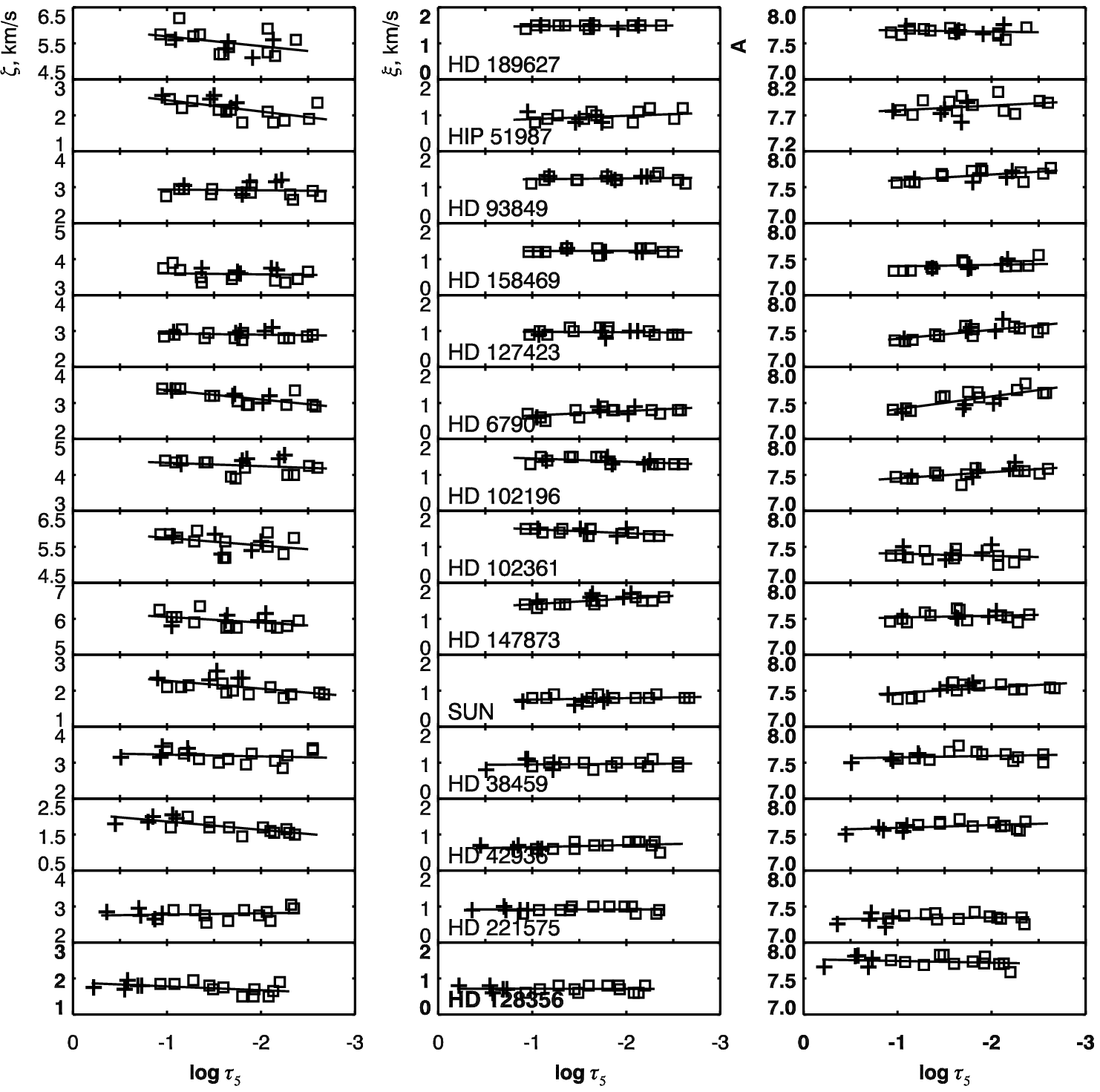}
     }
  %\caption
   \vspace {0.8 cm}
  {\small
Fig. 3. Obtained values of macroturbulence velocity $\zeta$, microturbulence velocity $\xi$, and iron abundance
$A = 12+ \log (N_{\rm Fe}/N_{\rm H})$ as functions of mean depth $\log \tau_5$ of spectral-line formation.
The data for Fe I and Fe II lines are denoted by squares and plus signs. Solid straight lines represent
the results of linear approximation.
 }
\label{prof1}
 \end{figure}

%%%%%%%%%%%%%%%%%%%%%%%%%%%%%%%%%%\lhd%%%%%%%%%%%%%%%%%%
%%==========================================================

%___________________________________ Table 2
\begin{table}
\centering
 \caption{\small Parameters of the studied stars and the values of micro- and 
 macroturbulence velocities, rotation velocity,
and iron abundance determined via Fourier analysis and averaged over all lines.  }
 \vspace {0.3 cm}
\label{T:2}

\footnotesize
\begin{tabular}{lcccccccccc}
\hline\hline
Star &  $T_{\rm eff}$&$\log g$&[M/H]&$M/M_\odot$&Age&$\xi$& $\zeta$& $v\sin i$& $A$  \\
       &   (K)  &               &     &  \cite{2019A&A...621A.112P}       & $10^9, \cite{2019A&A...621A.112P} $  & (km/s)      &  (km/s)     &  (km/s)     &      \\
\hline
HD 189627  &6210   &4.40   &~~0.07   &1.244& 4.0&1.48$\pm$0.04  &5.52$\pm$0.30  &5.93$\pm$0.02  & 7.67$\pm$0.05 \\
HIP 51987  &6158   &5.00   &~~0.27   &1.087& 7.2&0.96$\pm$0.14  &2.20$\pm$0.25  &2.09$\pm$0.03  & 7.81$\pm$0.10 \\
HD 93849   &6153   &4.21   &~~0.08   &1.268& 3.5&1.24$\pm$0.08  &2.92$\pm$0.16  &3.05$\pm$0.03  & 7.66$\pm$0.07 \\
HD 158469  &6105   &4.19   &$-$0.14  &1.223& 2.0&1.22$\pm$0.06  &3.61$\pm$0.14  &3.10$\pm$0.02  & 7.41$\pm$0.06 \\
HD 127423  &6020   &4.26   &$-$0.09  &1.107& 3.1&0.97$\pm$0.08  &2.90$\pm$0.10  &2.53$\pm$0.03  & 7.48$\pm$0.09 \\
HD 6790    &6012   &4.40   &$-$0.06  &1.089& 3.5&0.75$\pm$0.12  &3.16$\pm$0.18  &2.94$\pm$0.03  & 7.55$\pm$0.12 \\
HD 102196  &6012   &3.90   &$-$0.05  &1.395& 3.0&1.39$\pm$0.09  &4.26$\pm$0.19  &3.56$\pm$0.03  & 7.52$\pm$0.07 \\
HD 102361  &5978   &4.12   &$-$0.15  &1.250& 2.0&1.42$\pm$0.08  &5.62$\pm$0.25  &5.03$\pm$0.02  & 7.39$\pm$0.07 \\
HD 147873  &5972   &3.90   &$-$0.09  &1.493& 2.6&1.50$\pm$0.11  &5.95$\pm$0.17  &6.51$\pm$0.05  & 7.53$\pm$0.06 \\
Sun~~~~~~  &5777   &4.44   &~~0.00   &1.000& 4.6&0.78$\pm$0.08  &2.11$\pm$0.21  &1.84$\pm$0.02  & 7.52$\pm$0.07 \\
HD 38459   &5233   &4.43   &~~0.06   &0.882& 9.0&0.96$\pm$0.10  &3.20$\pm$0.17  &1.85$\pm$0.05  & 7.58$\pm$0.06 \\
HD 42936   &5126   &4.44   &~~0.19   &0.881&12.0&0.68$\pm$0.09  &1.74$\pm$0.18  &0.97$\pm$0.03  & 7.61$\pm$0.05 \\
HD 221575  &5037   &4.49   &$-$0.11  &0.823& 6.0&0.92$\pm$0.07  &2.79$\pm$0.14  &1.89$\pm$0.03  & 7.34$\pm$0.06 \\
HD 128356  &4875   &4.58   &~~0.34   &0.824&15.5&0.71$\pm$0.08  &1.74$\pm$0.14  &1.01$\pm$0.05  & 7.73$\pm$0.07 \\
\hline
\end{tabular}
\end{table}
\noindent

%%%%%%%%%%%%%%%%%%%%%%%%%%%%%%%%
\section{Discussion}
%%%%%%%%%%%%%%%%%%%%%%%%%%%%%%%%

{\bf Reliability of results.}
The reliability in determination of turbulence velocities  $\zeta$  and $\xi$, projection $V \sin i$ of 
the rotation velocity, and iron abundance A was verified for each line by directly comparing the calculated 
line profiles to observational data and finding the minimum deviation between them. It follows from Fig. 2 
that the shapes of the mean deviation profile for almost all stars are similar. The observed profile is 
narrower at the center and has broader wings than the Gaussian profile. This actually validates the conclusion 
made by Gray regarding the deviation of line profiles from the Gaussian (bell) shape. Our data suggest that 
these deviations are small: they average to below 0.25\% in the wings for all stars. The deviations for 
individual lines are as high as 1.5\% for certain stars. This may be attributed to the presence of weak 
invisible blends or a slight asymmetry of the observed profiles (especially those corresponding to the 
Sun and cooler K stars). It also follows from Fig. 2 that the observed profiles are deeper at the line 
centers than the calculated profiles. This is true for the entire studied sample with the exception of 
three stars with the highest macroturbulence and rotation velocities, which exceed 5 km/s. The broader 
the lines are ($V \sin i>3$, $\zeta>3$ km/s), the more accurate reproduction of profile broadening is 
provided by the isotropic Gaussian macroturbulence distribution. We have also calculated line profiles 
with the GM macroturbulence distribution and the RTM distribution with integration over the disk and 
concluded that RTM provides only a slight improvement; the shape of the deviation profile remains hardly 
changed. Since the average deviations from observations remain within the accuracy of the present analysis,
it is fair to assume that the obtained results are reliable.

Figure 3 presents the obtained values of macroturbulence velocity $\zeta$, microturbulence velocity $\xi$, 
and iron abundance $A = 12 + \log(N_{\rm Fe}{/}N_{\rm H})$ as functions of mean line formation depth 
$\log \tau_5$. The results for Fe I and Fe II lines do not feature any significant differences within the used 
MARCS atmospheric models. The results for each star were approximated with a linear dependence. Macroturbulence 
velocities have the maximum spread of values. A number of reasons for this may be suggested. The accuracy of 
Fourier analysis may be limited by the probable cross interference between $V \sin i$ and $\zeta$ in the 
comparison of residual transforms derived from the velocity model and observations. Weak anticorrelation 
between $\zeta$ and  $\xi$ may be observed (it is noticeable for certain stars in Fig.~3). In addition, 
varying influences of macroturbulence and rotation may shape almost the same velocity profile. The large 
spread is also attributable to the fact that macroturbulence may depend both on the convective driving force 
and on other factors (e.g., magnetic field or other features of stellar activity that are neglected in the 
present analysis). Despite the mentioned drawbacks of the method, we managed to obtain reliable results by 
minimizing the deviation between model and observational data for a large number of lines.

The values of $\zeta$, $\xi$, $V \sin i$, and $A$ averaged over all lines and their RMS deviations for each 
star are presented in Table 2 and Fig. 4. The RMS deviations are indicative of reliability of calculations 
(if there are no systematic variations of parameters with line intensity). These deviations were 
0.10--0.32~km/s for $\zeta$, 0.05--0.12 km/s for $\xi$, 0.07--0.14 for $A$, and 0.02--0.05 km/s for $V \sin i$. 
The deviations for macroturbulence were the largest, since $\zeta$ depends on depth within the photosphere.

For the Sun as a star the obtained values of iron abundance $A = 7.52 \pm 0.07$~dex, which agrees fairly well 
with the available data for the disk center ($7.47 \pm 0.04$~dex \cite{2015A&A...573A..26S}), and projection 
$V \sin i = 1.84 \pm 0.02$~km/s, which matches the synodic rotation velocity of the Sun at the equator 
(1.84~km/s \cite{2018ApJ...857..139G}), are indicative of reliability of the results of Fourier analysis.

{\bf Variation of the obtained parameters with depth in the photosphere.}
Our results suggest that macroturbulence velocity $\zeta$ varies with depth for most stars (Fig. 3). 
The most pronounced $\zeta$ variations correspond to stars with higher effective temperatures and 
stronger gravity (i.e., in hotter, denser, and less extensive atmospheres). Being an indicator of 
convection in stars, macroturbulence is related to the velocities of convective flows in subphotospheric 
and photospheric layers. The higher the convective velocities and the photosphere density, the steeper 
the gradient of $\zeta$. Therefore, stars with a more intense convection feature stronger macroturbulence, 
and its variations with depth are more pronounced. The values of $\zeta$ for the Sun as a star also clearly 
increases with depth. This has long been known from the studies of line profiles on the resolved solar 
disk \cite{1986SoPh..106..237G, 1982SoPh...78...39K, 2017PASJ...69...46T}.
%%%%%%%%%%%%%%%%%%%%%%%%%%%%%%%%%%%%%%%%%%% Figure 4
\begin{figure}[t]
 \centerline{
 \includegraphics   [scale=0.95]{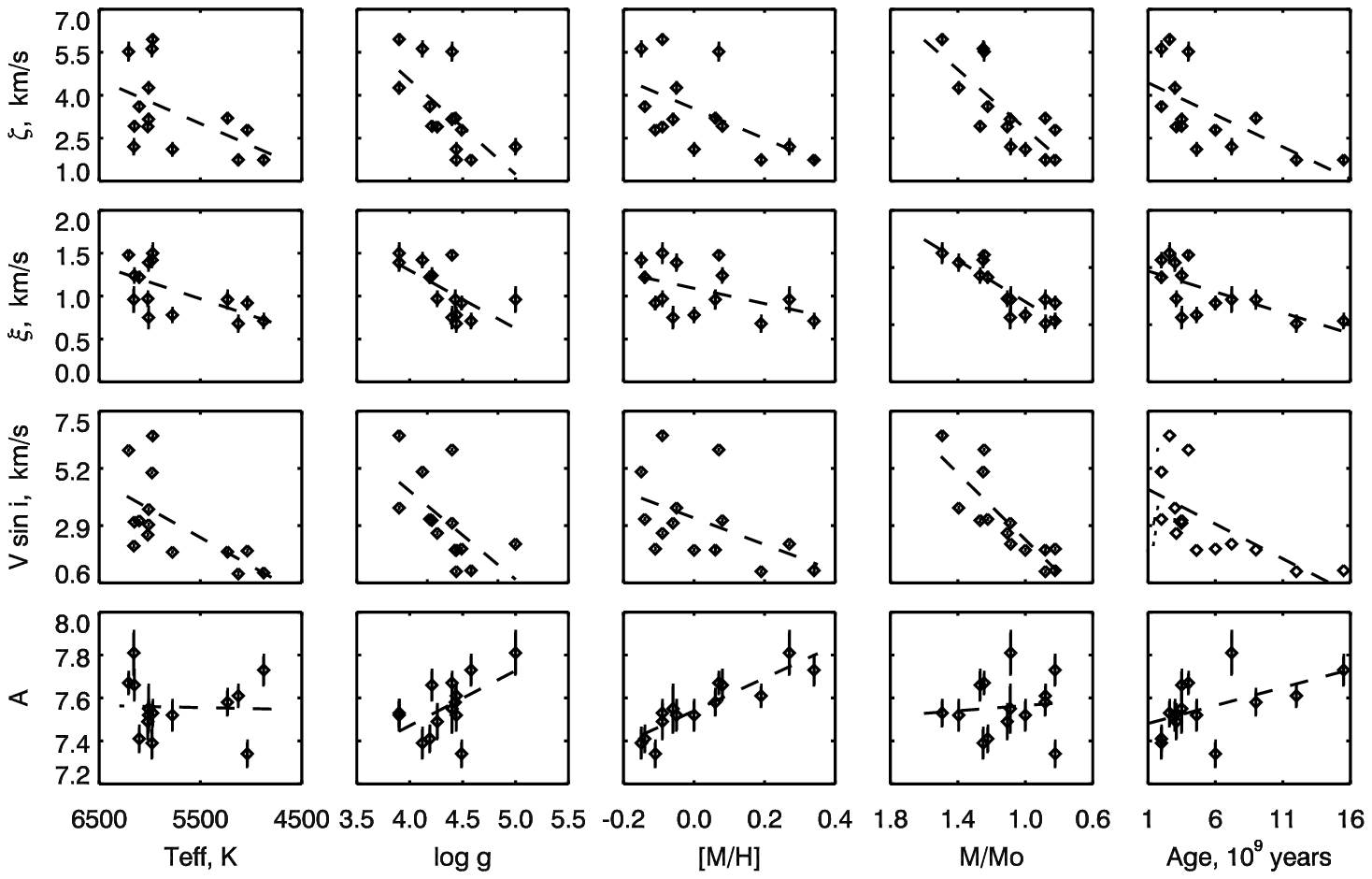}
 }
 % \caption
   \vspace {0.50 cm}
{\small
 Fig. 4. Average values of macroturbulence velocity $\zeta$, microturbulence velocity $\xi$, rotation
 velocity $V \sin i$,  and iron abundance $A$ as functions of the key stellar parameters: effective
 temperature $T_{\rm eff}$, surface gravity  $\log g$, metallicity [M/H], mass $M/M_\odot$, and age.
  }
\label{prof1}
 \end{figure}
%%%%%%%%%%%%%%%%%%%%%%%%%%%%%%%%%%%%%%%%%%%

It follows from Fig.~3 that microturbulence velocity $\xi$ varies only slightly with depth. It is fair 
to say that the gradients of microturbulence velocities for solar-type stars are insignificant. 
The value of $\xi$ increases with depth for certain stars, decreases in another group of stars, 
and remains almost constant in the third group. The microturbulence velocity for the Sun as a star 
varies little with depth; according to our data, its mean value is $0.8\pm0.1$. According to the results 
of analysis for the center of the solar disk \cite{1976AAfz...30...14G, 1986SoPh..106..237G, 1982SoPh...78...39K}, 
$\xi$ increases in deep photospheric layers and decreases with depth above the temperature minimum; its mean 
values for the center and the limb of the disk are $\xi=0.8$--1.0 and 1.4--1.7~km/s. It is likely 
that the effect of averaging over the disk masked the variation of $\xi$ with depth. 
Gray \cite{1978SoPh...59..193G} has pointed out that the data on variation of turbulence velocities with 
depth remain contradictory and do not allow one to draw definite conclusions.

Iron abundance $A$ should not vary from one line to another and should not depend on depth in the
photosphere. It follows from Fig. 3 that the values of $A$ for certain stars and the Sun decrease with depth, 
while other stars do not manifest such a dependence. It may be noted that $\xi$ and $A$ are not anticorrelated 
and that the disregard for non-LTE effects is not the cause for variation of $A$ with depth, since the values of 
$A$ obtained for Fe I and Fe II lines agree with each other. The variation of iron abundance with depth is 
likely to be related to errors in determination of equivalent widths of the observed weak lines, which may 
be underestimated due to inaccurate setting of the continuum level. The equivalent widths of strong lines
 may be overestimated due to the presence of invisible and unaccounted blends in broader line wings.

The obtained values of rotation velocity $V \sin i$ reveal hardly any variation from one line to the other 
in all stars; therefore, they are not shown in Fig.~3.

{\bf Variation of the obtained parameters along the HR diagram.}
The studied stars may be divided into two groups with effective temperatures of $\approx$6000 and 
$\approx$5000~K. Therefore, it is hard to identify a dependence on $T_{\rm eff}$ along the HR diagram. 
Figure 4 shows the values of $\zeta$, $\xi$, $V \sin i$,  and $A$ averaged over all lines for each star. 
The obtained turbulence velocities vary by a factor of two on average from the hottest stars to the coolest 
ones. In general, $\zeta$, $\xi$, and $V \sin i$ increase with temperature $T_{\rm eff}$ and mass $M/M_\odot$ 
but become smaller as surface gravity $\log g$, metallicity [M/H], and the age of a star increase. It follows 
from Fig.~4 that the variation of $V \sin i$ with age is the most pronounced; the age dependences of $\zeta$ 
and $\xi$ are weaker. These variation patterns agree with the ones determined earlier.

%%%%%%%%%%%%%%%%%%%%%%%%%%%%%%%%%%%%%%%%%%% Figure 5

\begin{figure}[t]
 \centerline{
 \includegraphics   [scale=0.95]{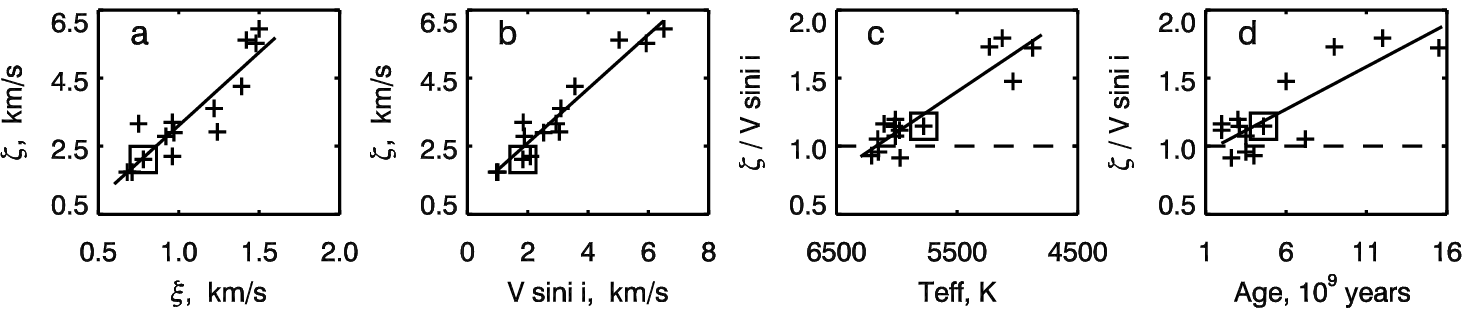}
     }
  %\caption
   \vspace {0.5 cm}
  {\small
Fig. 5. Dependences of the average values of macroturbulence velocity $\zeta$ on microturbulence
velocity $\xi$  and rotation velocity $V \sin i$ and dependences of ratio $\zeta/V \sin i$ on
effective temperature $T_{\rm eff}$ and the age of a star.
Squares represent data for the Sun.
 }
\label{prof1}
 \end{figure}

%%%%%%%%%%%%%%%%%%%%%%%%%%%%%%%%%%%%%%%%%%%

Figure 5 reveals an almost linear relationship between micro- and macroturbulence velocities: 
$\zeta \approx 4.21\xi -1.11 $. This is the reason why their dependencies on stellar parameters are similar. 
The macroturbulence velocity decreases from 4~km/s for hot stars to 2~km/s for cool stars; its solar value 
is 2.1~km/s. According to \cite{1982ApJ...255..200G}, the mean convective velocity decreases from 5.3~km/s 
in F5 V stars to a near-zero value in G8 V stars, but it increases again in even cooler stars. The convective 
velocity for the Sun is $1.9\pm0.2$~km/s. The obtained values of $\zeta$ for solar-type stars do not contradict 
the conclusions made in \cite{1982ApJ...255..200G} regarding the proportionality between $\zeta$ and convective 
velocities.

%%%%%%%%%%%%%%%%%%%%%%%%%%%%%%%%%%%%%%%%%%% Figure6
 \begin{figure}[t]
 \centerline{
% {\includegraphics   [scale=0.75]{fig3_new_5412.eps}}
 \includegraphics   [scale=0.8]{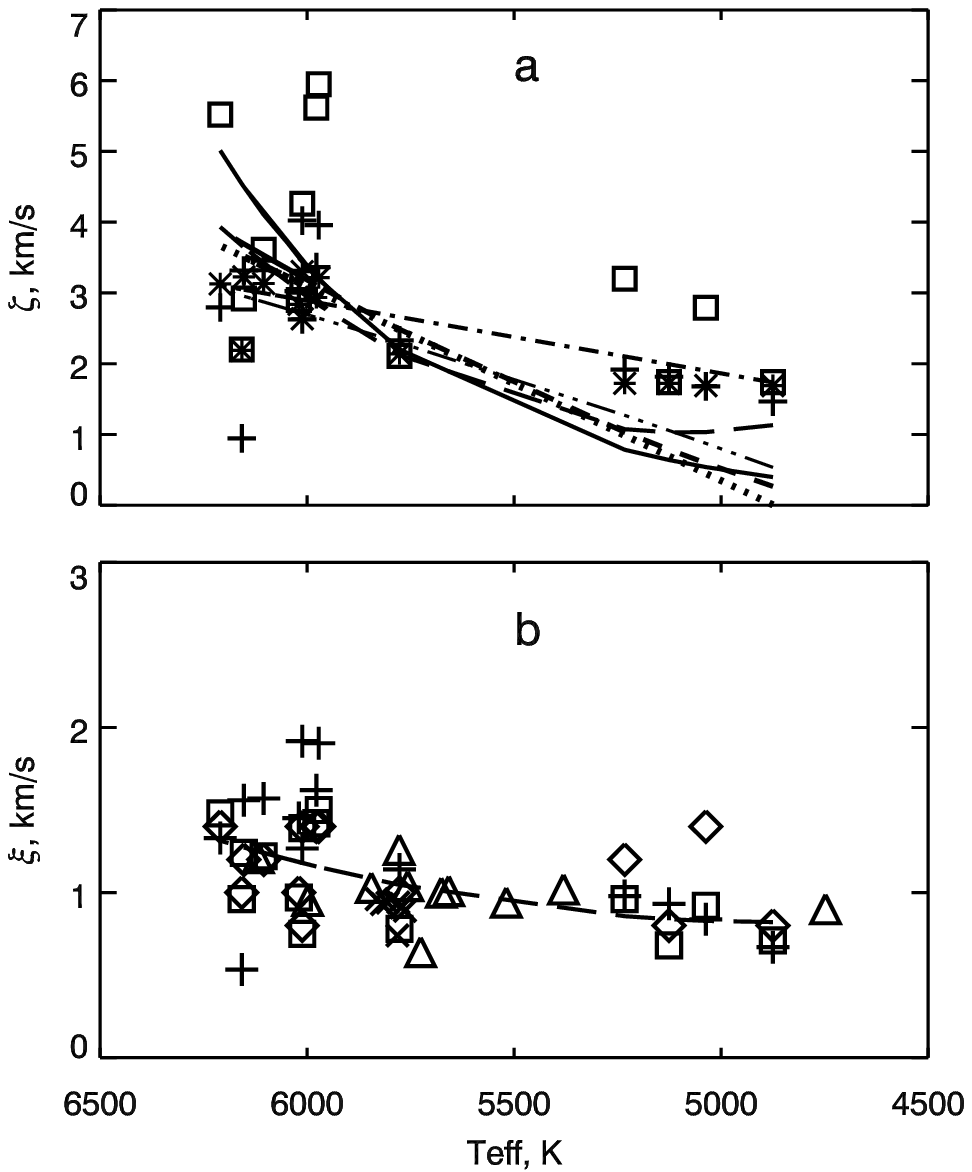}
    }
  %\caption
%   \vspace {0.5 cm}
  {\small
Fig. 6. Dependences of (a) macroturbulence velocity $\zeta$ (squares correspond to data from this study; 
asterisks, Doyle et al. \cite{2014MNRAS.444.3592D}; 
plus signs, Gonzalez \cite{1998A&A...334..221G}; 
dotted curve, Gray \cite{1984ApJ...281..719G}; 
dash-and-dot curve, Saar \& Osten \cite{1997MNRAS.284..803S}; 
dash-and-dot curve, Valenti \& Fischer \cite{2005ApJS..159..141V}; 
dashed curve, Bruntt et al. \cite{2010MNRAS.405.1907B}; 
solid curve, Brewer et al. \cite{2016ApJS..225...32B}; 
long dashed curve,  Takeda \& UeNo \cite{2017PASJ...69...46T})
and (b) microturbulence velocity $\xi$ (squares correspond to data from this study; 
plus signs, Nissen \cite{1981A&A....97..145N}; 
triangles, Neves et al. \cite{2009A&A...497..563N}; 
crosses, Sousa  et al. \cite{2011A&A...533A.141S};
diamonds, Ivanyuk et al. \cite{2017MNRAS.468.4151I}; 
long dashed curve, Bruntt et al. \cite{2010MNRAS.405.1907B})
 on the effective temperature. All velocities $\zeta$ are reduced to the
scale of macroturbulence velocities with the Gaussian distribution ($\zeta_G$).
 }
\label{prof1}
 \end{figure}

%%%%%%%%%%%%%%%%%%%%%%%%%%%%%%%%%%%%%%%%%%%

The spread of $\zeta$ and $\xi$ values in stars with equal effective temperatures is largely due to the 
differences in $\log g$. Turbulence velocities decrease as surface gravity $\log g$ increases, since the 
photosphere becomes denser and thinner. The chemical composition of the photosphere (or metallicity) may 
also affect the temperature dependence. The higher the concentration of metals, the more opaque and denser 
the photosphere; as a result, turbulence gets weaker. A well-marked dependence of turbulence velocities 
on $T_{\rm eff}$ in FGK stars may be obtained only if the studied stars would have the same values 
of $\log g$ and metallicity.

The results of our spectrometric measurements of $V \sin i$ as a function of stellar parameters are presented 
in Fig.~4. It can be seen that the trends in variation of $V \sin i$ and $\zeta$ are similar. Apparently, this 
is a manifestation of the relation between macroturbulence and rotation through convection. Macroturbulence 
depends on the properties of convection, which is affected by stellar rotation. This rotation is the driving 
force behind magnetic activity in convective layers. Emerging magnetic fields give rise to the chromospheric 
activity. The reduction in rotation velocity at lower $T_{\rm eff}$ is usually attributed to the deceleration 
of rotation induced by mass ejections (stellar wind). The stellar wind of main-sequence F0 stars (and cooler 
ones) may be driven by convection and chromospheric activity. Thus, rotation has an effect on convection, and 
convection affects macroturbulence and, via the magnetic field and chromospheric activity, rotation. This 
validates the relationship between macroturbulence and rotation (Fig. 5), which may be
represented by the following empirical formula: $\zeta \approx 1.25+0.68*V \sin i$. It is also instructive 
to trace the variation of ratio $\zeta {/} V \sin i$  with $T_{\rm eff}$ (Fig.~5). It can be seen that the 
average value of this ratio is 1 for hot stars and 1.7 for cooler K stars. Therefore, the macroturbulence 
velocity for solar-type stars decreases slower with age than the rotation velocity.

%___________________________________ Table 3
\begin{table}
\centering
% \caption{
 \small  Table 3. Results of different studies focused on determining the parameters
 of turbulence velocities for the Sun as a
star. Asterisks denote the values obtained by converting $\zeta_{\rm RT}$ to $\zeta_{\rm G}$ and vice versa.
% } 
\vspace {0.3 cm}
\label{T:1}
\footnotesize{
%\small{
\begin{tabular}{llllll}
\hline
$\zeta_{\rm RT}, km/s$ & $\zeta_{\rm G}$, km/s & $\xi$, km/s   & $V \sin i$, km/s & Method & Reference \\
\hline
 \multicolumn{6}{c} {Very strong lines }  \\
2.3               &  1.5$^*$          &0.5    & 2.0   & Line profile & \cite{1995PASJ...47..337T}\\
 \multicolumn{6}{c} {Strong lines}\\
3.0$^*$             &  2.0            &0.5       & 1.9 & Line profile &  \cite{2017PASJ...69...46T}\\
3.1               &  2.15$^*$         &0.5      & 1.9     & Fourier transform  & \cite{1977ApJ...218..530G}\\
2.58$^*$         &  1.72         &1.07    & 1.9  & Line profile& \cite{1990SvA....34..260G}\\
2.6               &  1.9            &0.70     & 1.85   & Fourier transform  &  \cite{1998KPCB...14..169S}\\
2.89             &  1.99          &0.85     & 1.75     & Fourier transform  &  \cite{2017KPCB...33..217S}\\
2.92$^*$           &  2.03          &0.78     & 1.84      & Fourier transform   & This study\\
 \multicolumn{6}{c} {Weak lines}  \\
3.45$^*$          &  2.3               &1.2    & 2.03     &  Line profile&\cite{1984BSolD...8...70S}\\
3.7               &  2.5$^*$           &0.5      & 2.0    & Line profile&  \cite{1995PASJ...47..337T}\\
3.45$^*$          &  2.3               &0.8      & 1.9   &  Line profile& \cite{1997MNRAS.284..803S}\\
3.45$^*$          &  2.3               &0.5      & 1.9  & Line profile&  \cite{2017PASJ...69...46T}\\
3.8               &  2.64$^*$         &0.5       & 1.9    &Fourier transform   &  \cite{1977ApJ...218..530G}\\
3.77              &  2.6$^*$          & --        & 1.75  & Fourier transform   &  \cite{2018ApJ...857..139G}\\
3.22              &  2.22           &0.85     & 1.75  & Fourier transform   &  \cite{2017KPCB...33..217S}\\
3.15$^*$          &  2.19           &0.78     & 1.84      & Fourier transform   & This study\\
 \multicolumn{6}{c} {Strong and weak lines }\\
3.2              &  2.13$^*$         & --     & 2.2    & Fourier transform  &  \cite{1978ApJ...224..584S}\\
3.21              &  2.14$^*$         &0.85    & 1.9  & Line profile&  \cite{2014MNRAS.444.3592D}\\
3.5                &  2.2$^*$           &0.40  & 1.7  & Fourier transform   &  \cite{1998A&A...334..221G}\\
\hline
\end{tabular}
}
\end{table}
\noindent
%%%%%%%%%%%%%%%%%%%%%%%%%%%%%%%%%%\

%%%%%%%%%%%%%%%%%%%%%%%%%%%%%%%%
\section{Comparison with the results of other studies }
%%%%%%%%%%%%%%%%%%%%%%%%%%%%%%%%

The macroturbulence velocity was determined in different studies with either the isotropic Gaussian
model ($\zeta_{\rm G}$) or the radial-tangential model ($\zeta_{\rm RT} $). The ratio between 
$\zeta_{\rm RT} $ and $\zeta_{\rm G}$ values was estimated at $\zeta_{\rm RT}{/}\zeta_{\rm G} = 1.44 $ 
in \cite{1978SoPh...59..193G} and $\approx$1.5 in \cite{1978SoPh...59..193G} by fitting the Fourier transforms. 
Takeda \cite{2017PASJ...69...46T} 
suggests that this ratio may depend on the additional line broadening. He obtained $\approx 1.67$ by fitting 
line profiles for the solar flux. We have also determined $\zeta_{\rm RT} {/} \zeta_{\rm G}$ by fitting the 
observed and calculated line profiles with RTM and GM and obtained a value of $\approx$1.5, which was used to convert 
$\zeta_{\rm RT} $  into $\zeta_{\rm G}$ derived from line profiles. The value of 
$\zeta_{\rm RT} {/} \zeta_{\rm G} = 1.44$ \cite{1978SoPh...59..193G} was used to convert the $\zeta_{\rm RT} $  
estimates derived from Fourier transforms.

Figure 6a shows the values of macroturbulence velocities in FGK stars determined in 
\cite{2016ApJS..225...32B, 2010MNRAS.405.1907B, 2014MNRAS.444.3592D, 
1998A&A...334..221G,1984ApJ...281..719G, 1997MNRAS.284..803S, 
2017PASJ...69...46T, 2005ApJS..159..141V} as functions of the effective
 temperature. It should be noted that the coincidence between all empirical
curves after the convertation $\zeta_{\rm RT} $ to $\zeta_{\rm G}$ is considerably better than that in a 
similar plot  of Takeda \cite{2017PASJ...69...46T}. Figure 6b presents the estimates of microturbulence velocities 
obtained in \cite{2010MNRAS.405.1907B, 2017MNRAS.468.4151I, 
2009A&A...497..563N, 1981A&A....97..145N,2011A&A...533A.141S}. Our 
estimates (squares) agree fairly well with the results of other studies, verify the reliability of our 
analysis, and lend credibility to the dependences revealed earlier.

The following facts regarding macroturbulence are already known \cite{1978SoPh...59..193G, 2005oasp.book.....G}. 
Macroturbulence velocity $\zeta$ is a steeper function of the spectral type than microturbulence velocity $\xi$. 
Velocity $\zeta$ decreases rapidly toward later spectral types from F0 to K0. The microturbulence velocity 
decreases with a large spread of values in the hotter region from A5 to G0 and then increases somewhat to K stars. 
The weaker the gravity on the stellar surface, the higher the turbulence velocities. The rotation velocity increases 
markedly along the main sequence from F stars (50--5~km/s) to B0--A0 stars ($ \approx$200~km/s) and reaches 
the measurement limit for cool GK stars. Stars with lower luminosities have higher rotation velocities than 
main-sequence stars. All this is confirmed by the data on micro- and macroturbulence and rotation of solar-type 
stars obtained in the present study.

{\bf Estimates for the Sun.}
The Sun is a reference for studies of other stars, and the solar flux spectrum is often used to test the
results. The recent data for the Sun as a star are presented in Table 3. The large spread of 
$\zeta_{\rm RT}$ and $\zeta_{\rm G}$ values is attributable primarily to the variation of macroturbulence 
with depth in the solar photosphere. The values of $\zeta_{\rm G}$ determined based on weak and strong lines 
are 2.2--2.6 and 1.9--2.0~km/s, respectively. The microturbulence velocity value falls within the range from 0.4 
to 1.2 km/s and is independent of the line intensity. The most likely causes of discrepancies between the 
$\xi$  estimates are the errors in determination of the observed equivalent widths, oscillator 
strengths, and the damping constant. Therefore, a large spread of turbulence velocity values is to be expected 
in the analysis of stellar spectra.

{\bf Comparison between our data and the estimates from  \cite{2017MNRAS.468.4151I}}.
It is instructive to compare our data to the estimates obtained for the same stars. The spectra of
107 solar-type stars taken from a set of high-quality homogeneous observational data Jenkins et al. 
\cite{2009MNRAS.398..911J}  were analyzed by A. Ivanyuk et al. \cite{2017MNRAS.468.4151I}, and the 
effective temperature, the surface gravity, and the chemical composition of these stars were determined. 
In addition, the microturbulence velocity and the rotation parameter were estimated by analyzing the 
line profiles with a constant macroturbulence velocity $\zeta_{\rm G} = 2$~km/s assumed for all stars. 
This assumption may introduce a certain error into the obtained results, since macroturbulence depends 
on $T_{\rm eff}$ and $\log g$. Figure~7 shows the correlative dependences of $\xi$,  $V \sin i$, and 
iron abundance $A$ estimated in the present study and in \cite{2017MNRAS.468.4151I}. As expected, the 
values of $V \sin i$ obtained in \cite{2017MNRAS.468.4151I} are higher, since they compensate for the 
lack of macroturbulent broadening. The data on microturbulence velocity and iron abundance generally 
agree within the limits of error.

%%%%%%%%%%%%%%%%%%%%%%%%%%%%%%%%%%%%%%%%%%%
%%%%%%%%%%%%%%%%%%%%%%%%%%%%%%%%%%%%%%%%%%% Figure 7
 \begin{figure}[t]
 \centerline{
% {\includegraphics   [scale=0.75]{fig3_new_5412.eps}}
 \includegraphics   [scale=0.8]{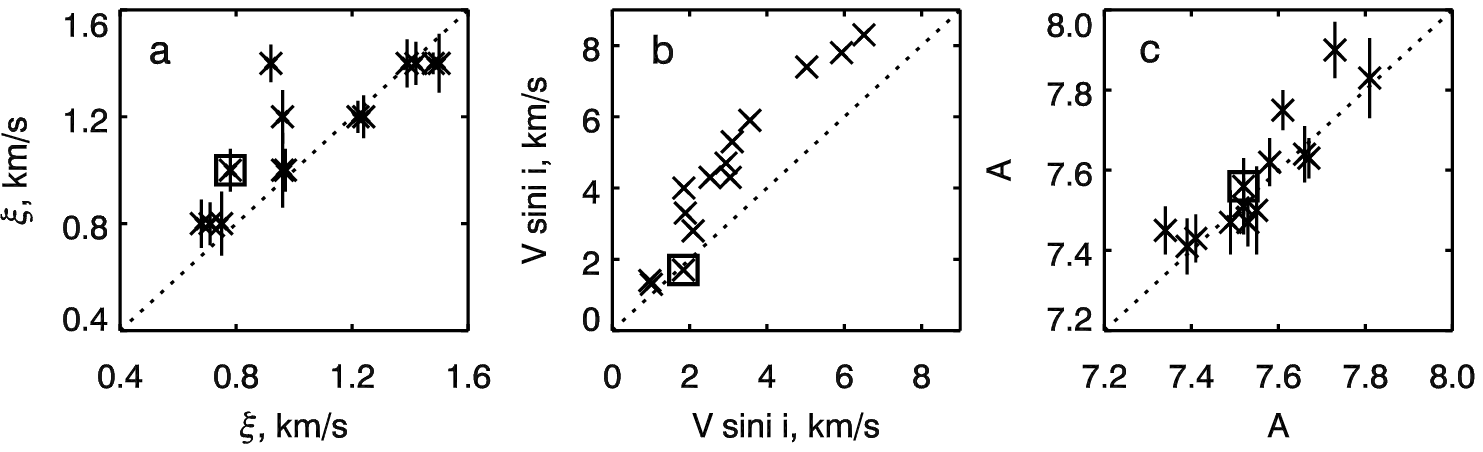}
    }
  %\caption {}
   \vspace {0.5 cm}
  {\small
Fig. 7. Correlative dependences of the mean values of microturbulence $\xi$, rotation $V \sin i$,
and iron abundance $A$ obtained in \cite{2017MNRAS.468.4151I} and in this study for the same stars.
Squares represent the data for the Sun; the dotted line is the line of equal values.
 }
\label{prof1}
 \end{figure}

%%%%%%%%%%%%%%%%%%%%%%%%%%%%%%%%%%%%%%%%%%%

\section{Conclusions }

The line-of-sight turbulence velocities, the rotation velocity, and the iron abundance of solar-type
stars were studied based on HARPS spectroscopic data. Fourier analysis with the isotropic Gaussian
model of micro- and macroturbulence was performed for 17 iron lines in the spectra of 13 stars and the
Sun. Since all properties of the atmosphere are defined by the energy flow and the gas density (or, in other
words, by temperature and the surface gravity), the obtained results were tested for dependence on these
parameters. The results of this test agreed in general with the dependences that were already known. We
have also tried to determine the variation of turbulence velocities with depth in stellar atmospheres. The
key findings are as follows.

The macroturbulence velocity in stellar atmospheres increases with effective temperature and with
depth in the photosphere. It decreases as the surface gravity gets higher. The gradient of its 
variation with depth becomes steeper as the effective temperature increases and the surface gravity 
gets stronger. The macroturbulence velocity for the coolest stars is almost independent of depth.

The macroturbulence and microturbulence velocities are closely related. In general, microturbulence
intensifies together with macroturbulence in the atmospheres of solar-type stars. The dependences of
these velocities on the fundamental parameters are also similar; the only difference is that they are less
steep for microturbulence. The microturbulence velocity varies little with depth in the atmospheres of the
studied stars. It is almost constant for the Sun and several stars, while it either increases or decreases
slightly with depth in other groups of stars.

The dependences of the projected rotation velocity on the effective temperature and the surface gravity
are similar to those of the turbulence velocities. The higher the effective temperature and the lower the
surface gravity, the faster the axial rotation of a star. The greater the age and the smaller the mass of 
a star, the lower the rotation velocity.

The stellar rotation velocity is correlated with macroturbulence. The higher the rotation velocity, the
higher the macroturbulence velocity. The ratio between the macroturbulence and rotation velocities is
approximately equal to unity for stars with an effective temperature of $\approx$6000~K. This ratio for 
cooler stars with an effective temperature of $\approx$5000~K is 1.7. The age dependence of the rotation 
velocity is more pronounced than that of the velocity of macroturbulent motions.

{\bf Acknowledgments.}
I thank Ya. Pavlenko and A. Ivanyuk for providing stellar spectra and for fruitful discussions.

\vspace{0.3cm}
{\bf Funding.}
This study was funded as part of the routine financing program for institutes of the National Academy
of Sciences of Ukraine.

\end{document}